\documentclass[aps,pra,showpacs]{revtex4-1}
\usepackage{graphicx}
\usepackage{subcaption}
\usepackage{color}
\usepackage[font=footnotesize]{caption}

\usepackage{fullpage}
\begin{document}

\title{Redistribution of light frequency by multiple scattering in a resonant atomic vapor.}
\author{Jo\~ao Carlos de A. Carvalho}
\affiliation{Laborat\'orio de Espectroscopia \'Otica, Universidade Federal da Para\'iba - Caixa Postal 5086, 58051-900 Jo\~ao Pessoa, PB, Brazil}
\author{Hugo L. D. de Souza Cavalcante}
\affiliation{Departamento de Inform\'atica, Centro de Inform\'atica, Universidade Federal da Para\'iba, Avenida dos Escoteiros -s/n$^\circ$ Mangabeira VII, 58055-000 Jo\~ao Pessoa, PB, Brazil}
\author{Marcos Ori\'a}
\author{Martine Chevrollier}
\affiliation{Laborat\'orio de Espectroscopia \'Otica, Universidade Federal da Para\'iba - Caixa Postal 5086, 58051-900 Jo\~ao Pessoa, PB, Brazil}
\author{T. Passerat de Silans\email{thierry@otica.ufpb.br}}
\affiliation{Laborat\'orio de Superfícies - DF, Universidade Federal da Para\'iba - Caixa Postal 5086, 58051-900 Jo\~ao Pessoa, PB, Brazil}
\affiliation{UAF, Universidade Federal da Campina Grande, 58051-970 Campina Grande, PB, Brazil}
\begin{abstract}
The propagation of light in a resonant atomic vapor can \textit{a priori} be thought of as a multiple scattering process, in which each scattering event redistributes both the direction and the frequency of the photons. Particularly, the frequency redistribution may result in L\'evy flights of photons, directly affecting the transport properties of light in a resonant atomic vapor and turning this propagation into a superdifusion process. Here, we report on a Monte-Carlo simulation developed to study the evolution of the spectrum of the light in a resonant thermal vapor. We observe the gradual change of the spectrum and its convergence towards a regime of Complete Frequency Redistribution as the number of scattering events increases. We also analyse the probability density function of the step length of photons between emissions and reabsorptions in the vapor, which governs the statistics of the light diffusion. We observe two different regime in the light transport: superdiffusive when the vapor is excited near the line center and normal diffusion for excitation far from the line center. The regime of Complete Frequency Redistribution is not reached for excitation far from resonance even after many absorption/reemission cycles due to correlations between emitted and absorbed frequencies.
\end{abstract}
\maketitle

\section{Introduction}

Light scattering in atomic vapors has been studied for a long time, for resonant as well as for non-resonant radiation \cite{Compton,Milne}. When light is resonant with atomic transitions, many absorption-reemission cycles may occur before a photon leaves the vapor volume, a phenomenon known as imprisonment of resonant radiation \cite{Holstein47} or radiation trapping \cite{Molisch94}. This multiple scattering process alters the spatial distribution of excitation in a resonant vapor, as well as the frequency spectrum of the light, because photons usually suffer a frequency shift at each scattering event. The importance of the frequency redistribution for light diffused in resonant vapors was already acknowledged in the 1930s. Radiation trapping and frequency redistribution are long-studied topics in astrophysics, for instance in the analysis of the radiation emitted by nebulae \cite{Spitzer44,Unno51,Frisch79,Zanstra46}.\\

The broadening of spectral lines favors the emergence of long steps for photons in a resonant vapor \cite{Kenty32}, in such a way that the light escapes the vapor volume faster than expected in a diffusive description \cite{Zemansky27,Zemansky30}. Occasional very long steps of photons result in the impossibility of defining a mean square displacement. To overcome the failure of a diffusive model to describe light transport in resonant vapors, Holstein \cite{Holstein47} developed a Boltzmann-like integro-differential equation taking into account the strong spectral variations of the absorption coefficient around resonance frequencies. Holstein equation is usually solved for Complete Frequency Redistribution (CFR), i.e. assuming that there is no correlation between the frequency of the absorbed (incident) and of the reemitted photons. In this case, the probability density function (PDF) $\Theta(\nu)$ of the incident radiation is equal to the PDF $\Phi(\nu)$ of the vapor absorption \cite{Payne74,Hishikawa95}. CFR usually occurs when atomic collisions are very frequent and destroy the correlations between incident and scattered photons \cite{Frisch79}.\\

For low-density vapors, where interatomic collisions can be ignored, the frequency of the reemitted photon is exactly the same as the frequency of the absorbed photon, in the atom's rest frame (ARF), i.e. there is no frequency redistribution in the ARF. However, in the referential of the laboratory, the Doppler effect due to the motion of the scattering atoms shifts the frequency of the photons, leading to Partial Frequency Redistribution (PFR) in the laboratory frame. One can define a redistribution function $R(\nu',\nu)$ that gives the probability $R(\nu',\nu) d\nu' d\nu$ that a photon of frequency between $\nu$ and $\nu + d\nu$ is absorbed and that the reemitted photon has a frequency between $\nu'$ and $\nu' + d\nu'$. This redistribution function was calculated by Unno \cite{Unno51} for an infinitely sharp but Doppler-broadened line. Latter, Unno \cite{Unno52} and Hummer \cite{Hummer62} calculated the redistribution function for a line profile with both natural and Doppler broadening. As Holstein equation is not easy to solve in PFR conditions, the study of light transport in resonant scattering media is usually carried out through Monte-Carlo (MC) simulations \cite{Avery68}. Such numerical simulations allow to follow both the spatial distribution of the light excitation and the frequency spectrum of the light after a given number of scattering events. MC simulations have been used to obtain information such as the number of scattering events that a photon undergoes before it escapes from a vapor cell with a particular geometry \cite{Avery68}; test the local thermodynamic equilibrium assumed by Holstein \cite{Klots72}; study frequency diffusion when a cloud of cold atoms is illuminated by a laser light detuned up to a few natural widths from resonance \cite{Labeyrie05}.\\ 

Besides its applications in astrophysics, radiation trapping is also a phenomenon of great interest in atomic physics laboratories. Radiation trapping increases the effective lifetime of the excited population \cite{Kibble}. Radiation trapping diffuses light incoherently and contributes to ground state decoherence in experiments of Coherent Population Trapping (CPT) and Electromagnetically Induced Transparency (EIT) \cite{Matsko01}. The diffusion of photons in a vapor modifies the spatial distribution of excited atoms \cite{Heron} and thus plays a role in the spatial distribution of saturation \cite{Ackemann97}.\\

The strong dependence of the transport properties of light on the spectral characteristics of the atom-photon interaction has led to an increasing interest in studying the statistics of the random walk of photons in a resonant vapor. It has been theoretically predicted that the length of the steps between two scattering events does not follow the statistics of normal distributions \cite{Pereira04,Pereira07,Chevrollier}. Anomalous, superdiffusion of light in atomic vapors occurs due to long steps taken by reemitted photons whose frequency lays far in the aisles of the curve of the absorption probability distribution. Superdiffusion is then characterized by a PDF of step length $l$ that asymptotically follows a power law $P(l)\sim l^{-k}$ with $k<3$. The step-length distribution depends on the spectrum of the incident radiation as well as on the absorption profile of the scattering vapor \cite{Pereira07}. All the standard atomic spectral lineshapes (Doppler, Lorentzian and Voigt) result in such a superdiffusive behavior, characterized by rare, large steps known as L\'{e}vy flights. For Voigt incident and absorption spectra, for instance, the theoretical prediction yields \cite{Pereira04} $k=3/2$. While it is trivial to observe the exponentially decaying step-length distribution of (monochromatic) laser photons in a Voigt-broadened vapor, measuring the PDF of the photons step length in the CFR regime is not an easy task. To do so, one may, for instance, prepare an incident radiation with a Voigt-like profile, which was achieved in Ref. \cite{NatPhys} by submitting laser photons to a few frequency-redistributing scattering events in an auxiliary vapor cell prior to sending them to the measurement cell \cite{Mercadier13}.

Here we are interested in this ''preparation" of a Voigt spectrum for the radiation for studies of photons steps statistics. We implement a MC routine based on first principles, whose aim is to analyse the PFR in the few initial scattering events. In particular, we want to determine how many scattering events are needed to generate an emission spectrum similar to the absorption one, i.e., a CFR. From this simulation we can also obtain the evolution of the PDF of the photons step length. The knowledge of the incident spectral profile and of the PDF of the step size is crucial in the experimental study of the statistics of photon diffusion in a vapor \cite{NatPhys,Mercadier13}.

\section{Monte-Carlo simulations}
In this section we describe the Monte-Carlo simulation used to study the frequency redistribution of photons scattered in an atomic vapor.\\

We consider a photon in a resonant vapor and departing from the origin of the coordinate system with a detuning $\delta_I$ relative to the center of the atomic transition and with an isotropic random direction. This photon may be absorbed by atoms of the vapor and then reemitted many times. We call $i$ the scattering event number. After the $i_{th}$ scattering event the photon is reemitted isotropically, with a detuning $\delta_{i}$ related to the absorbed frequency and to the velocity of the emitting atom through Doppler effect. The photon travels a distance $l_i$ in the vapor before being absorbed again and the process of reabsorption/reemission is repeated until a given number of scattering events is reached. The frequency of the photon and the velocity of the emitting atom are recorded after each scattering event. Details of the Monte-Carlo simulation are given below.\\

The MC simulation is developed for two-level atoms with natural broadening of their excited level and with Doppler broadening. For such a system, the absorption profile is a Voigt profile with Voigt parameter $a=\frac{\Gamma}{ \Gamma_D}$, where $\Gamma$ is the natural line width, $\Gamma_D= 2 u/\lambda$ is the Doppler width at $1/e$ of its maximum, $u=\sqrt{2k_BT/m}$ is the half width of the Maxwell-Boltzmann velocity distribution at temperature $T$, $\lambda$ is the wavelength of the atomic transition, $m$ is the mass of the atom and $k_B$ is the Boltzmann constant. We use the value $a=0.01$ corresponding to the $D_2$ transition of $^{85}$Rb at room temperature ($T=294$ K). At such a temperature, the pressure broadening is negligible ($\sim 1$ kHz \cite{Pitchler}) and will not be taken into account.

\subsection{Photon step length}
The distance $l_i$ traveled by a photon between the $i_{th}$ emission and the $(i+1)_{th}$ absorption in the vapor is drawn from the Beer-Lambert probability distribution function, $P(l)dl=\alpha(\delta_i)\exp{\left(-\alpha(\delta_i)l\right)}dl$, which gives the probability that a photon of frequency $\delta_i$ be absorbed after having traveled a distance between $l$ and $l+dl$ see Refs.\cite{Klots72,Chevrollier}. $\alpha(\delta_i)$ is the absorption coefficient at detuning $\delta_i$ as measured in the laboratory frame. It is proportional to the Voigt profile and to the atomic vapor density.\\

\subsection{Velocity of the absorbing atoms}
\label{Velocity of the atom absorbing the photon}
\begin{figure}
        \centering
		\includegraphics[width=1.0\linewidth]{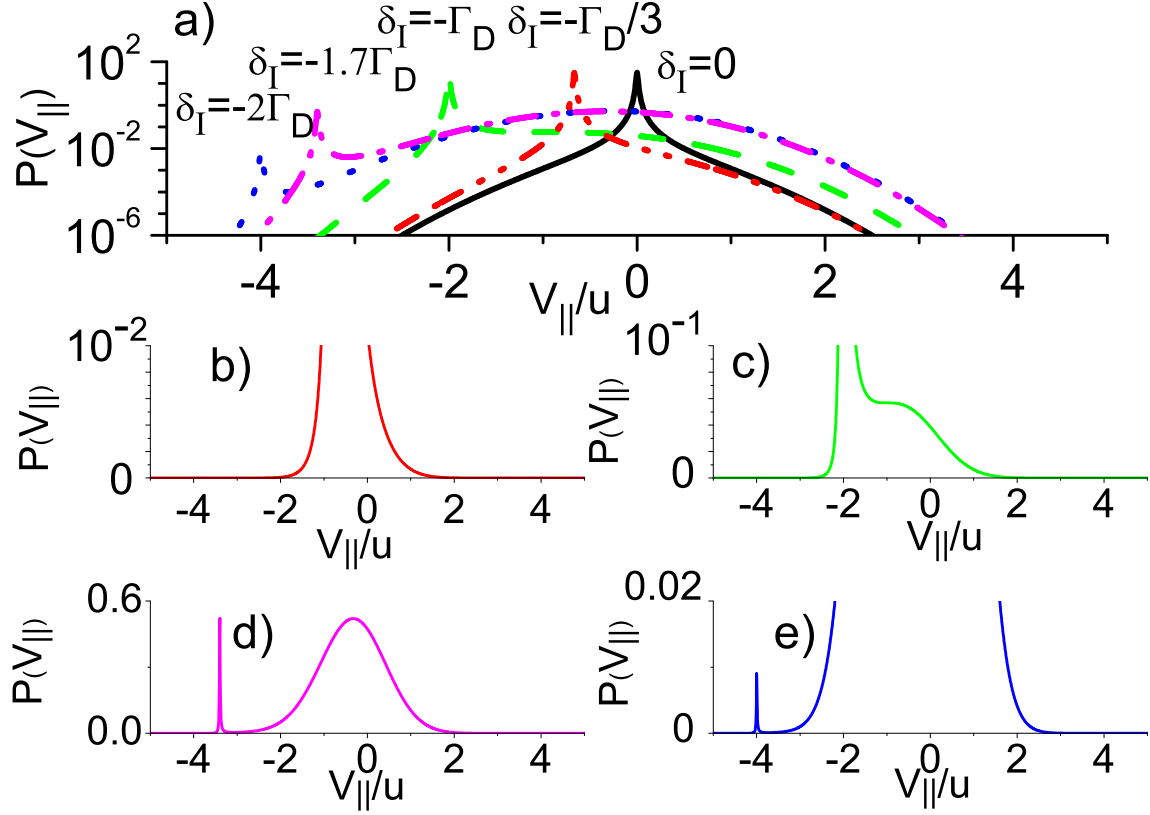}
        \caption{(Color online) Probability density function of the velocity component of absorbing atoms parallel to the direction of incident photons, for different excitation detunings $\delta_I$. The values of $V_{||}$ are normalized to the half width of the Maxwell-Boltzmann distribution at temperature of 294 K. (a) In log scale, (b-e) in linear scale (zoom). (b) for $\delta_I=-\frac{1}{3}\Gamma_D$ and (c) for $\delta_I=-\Gamma_D$: Photons with a small detuning are preferentially absorbed by atoms whose parallel velocity component Doppler compensates for the detuning. (d) for $\delta_I=-1.7\Gamma_D$ and (e) for $\delta_I=-2\Gamma_D$: very few atoms absorb the far-detuned photons at the line center in the atomic rest frame, see small narrow peaks. Most atoms absorb instead in the far-reaching wings of the Lorentz lineshape, where favorable MB probability compensates for the very weak absorption probability (this one decaying as $\delta^{-2}$ in the wings).}\label{fig:ProbV}
\end{figure}

An important variable in the determination of the frequency redistribution by radiation trapping is the velocity of the atoms that absorb and reemit the photons. The speed of the atoms in a vapor follows a Maxwell-Boltzmann (MB) distribution. However, for a given frequency profile of the incident radiation, some atoms will absorb the photons more favorably, depending on their velocity component in the photon direction. The probability $P_{\delta_i}(V_{||})$ that an atom with a parallel component of velocity $V_{||}$ (positive sign in the direction of the photon) absorb a photon with detuning $\delta_i$ is given by the product of the Doppler-shifted Lorentzian atomic lineshape, centered at $\lambda\delta_i$, and the Maxwell-Boltzmann distribution of $V_{||}$:
\begin{equation}
P_{\delta_i}(V_{||})\propto\frac{1}{1+4\delta^2_{A,i}/\Gamma^2}e^{-V^2_{||}/u^2}, \label{ProbVPar}
\end{equation}
where $\delta_{A,i}=\delta_i-(V_{||}/\lambda)$ is the detuning of the incoming photon in the atomic rest frame and $(V_{||}/\lambda)$ is the Doppler shift. The probability density function of $V_{||}$ of the absorbing atoms is shown in Figure \ref{fig:ProbV} for different detunings of the incoming photons.\\

The normal component of the atom velocity plays no role in the absorption process and is therefore independently drawn from a Maxwell-Boltzmann distribution at the temperature of the vapor.\\

The PDF of the parallel velocity component is composed of two physically distinct contributions (Equation \ref{ProbVPar}), with the wide MB distribution modulating the amplitude of the Doppler-shifted atomic Lorentzian lineshape. Due to this convolution of effects, the PDF of $V_{||}$ (Figure 1) evidences two qualitatively different situations: i) for an incident frequency close to the atomic resonance (small $\delta_i$), the Maxwell-Boltzmann probability of finding an atom with parallel velocity $V_{||}=\lambda\delta_i$ is high. The PDF of $V_{||}$ is a Lorentzian peak around $V_{||}=\lambda\delta_i$. This situation corresponds to an absorption at line center in the atomic rest frame ($\delta_{A,i}=0$, with a width $\Delta V_{||}\approx \lambda\Gamma$, see Figures \ref{fig:ProbV}(a) and Figures \ref{fig:ProbV}(b)(c)) for $\delta_I=-\frac{1}{3}\Gamma_D$ and $\delta_I=-\Gamma_D$, respectively. Note that for $\delta_I=-\Gamma_D$ (Figure \ref{ProbVPar}(c)) the photons' detuning is larger than the Doppler width and the MB probability of finding an atom with parallel velocity $V_{||}=\lambda\delta_I$ is one order of magnitude smaller than at resonance. The probability distribution of the parallel velocity component is still essentially a Lorentzian peak around $V_{||}=\lambda\delta_I$ ($P(0) < P(\lambda\delta_I)$) but atoms absorbing in the wings of this Lorentzian peak begin to give a noticeable contribution (broad peak at right of the narrow Lorentzian one). ii) For an incident frequency far from resonance, the probability of finding an atom with parallel velocity $V_{||}=\lambda\delta_i$ is very small, because of the very fast decay ($e^{-V^2_{||}}$) of the Maxwell-Boltzmann distribution. The photon is preferentially absorbed by an atom moving relatively slowly in the direction parallel to the incoming photon, i.e. inside the MB width but in the wings of the Lorentzian line-shape in the atomic rest frame (see Figures \ref{fig:ProbV}(a) and \ref{fig:ProbV}(e)). This situation occurs when $\frac{1}{1+4\delta_i^2/(a^2\Gamma_D^2)}>e^{-\delta_i^2/\Gamma_D^2}$, that is, when the probability $P(0)$ that the absorbing atom has $V_{||}=0$ is higher than the probability $P(\lambda\delta_i)$ that its parallel component of velocity is $V_{||}=\lambda\delta_i$.

We will call $\delta_L$ the detuning corresponding to the limit between those two behaviors for which $P(0) = P(\lambda\delta_I)$. For the parameter $a=0.01$ considered here, the limit between the two situations is around $\delta_i=\pm\delta_L=\pm1.7\Gamma_D$ (see Figures \ref{fig:ProbV}(a) and \ref{fig:ProbV} (d)). The two absorption mechanisms lead to different regimes of frequency redistribution of the reemitted photons.\\ 

\subsection{Emitted photon}
As mentioned before, we consider a low-pressure vapor where atomic collisions are not frequent. In such a situation, and considering the low intensity of the incident radiation \cite{Mollow69}, the scattering is elastic in the atomic rest frame, i.e. the emitted frequency is the same as the absorbed frequency \cite{Payne74}.\\

In the laboratory frame, the detuning $\delta_{i}$ of the emitted photon is Doppler-shifted, $\delta_i=\delta_{A,i}+\frac{\vec{n'}\cdot\vec{V}}{\lambda}$, where the direction of the emitted photon, of unit vector $\vec{n}'$, is drawn from an isotropic distribution. The Doppler shift ($\delta_{i}- \delta_{i-1}$) acquired in the $i_{th}$ emission constitutes the mechanism of frequency redistribution in the laboratory frame. If $\vec{n}$ is a unit vector in the direction of the incident photon, the atomic velocity component perpendicular to the plane ($\vec{n}, \vec{n'}$) does not play any role in the frequency change is this scattering event and the emitted frequency, as measured in the laboratory frame, is given by:
\begin{equation}
\delta_{i} = \delta_{i-1} + \frac{1}{\lambda} \left(V_\|(cos \theta -1)+ V_\bot sin \theta \right),
\label{eq:Dshift}
\end{equation}
where $V_\bot$ is the velocity component perpendicular to $\vec{n}$, in the ($\vec{n}, \vec{n'}$) plane and $\theta$ is the angle between $\vec{n}$ and $\vec{n'}$.
The Doppler frequency shift resulting from the isotropic reemission process therefore involves two of the atomic velocity components, $V_\|$ and $V_\bot$. 

\subsection{Boundaries of the system}

We consider an atomic vapor with no boundaries (we call it \textit{infinite vapor}). Simulating light transport in such an \textit{infinite vapor} helps us understand the frequency redistribution mechanism at each scattering event. Furthermore, we can observe the probability distribution of the photons step length in order to get a deeper insight into the statistically anomalous properties of radiation trapping.\\

\section{Results}
\subsection{Frequency redistribution}
\label{sec:freqredis}
\subsubsection{First scattering event}
\label{sec:freqredis1}
\begin{figure}
	\centering
		\includegraphics[width=1.0\linewidth]{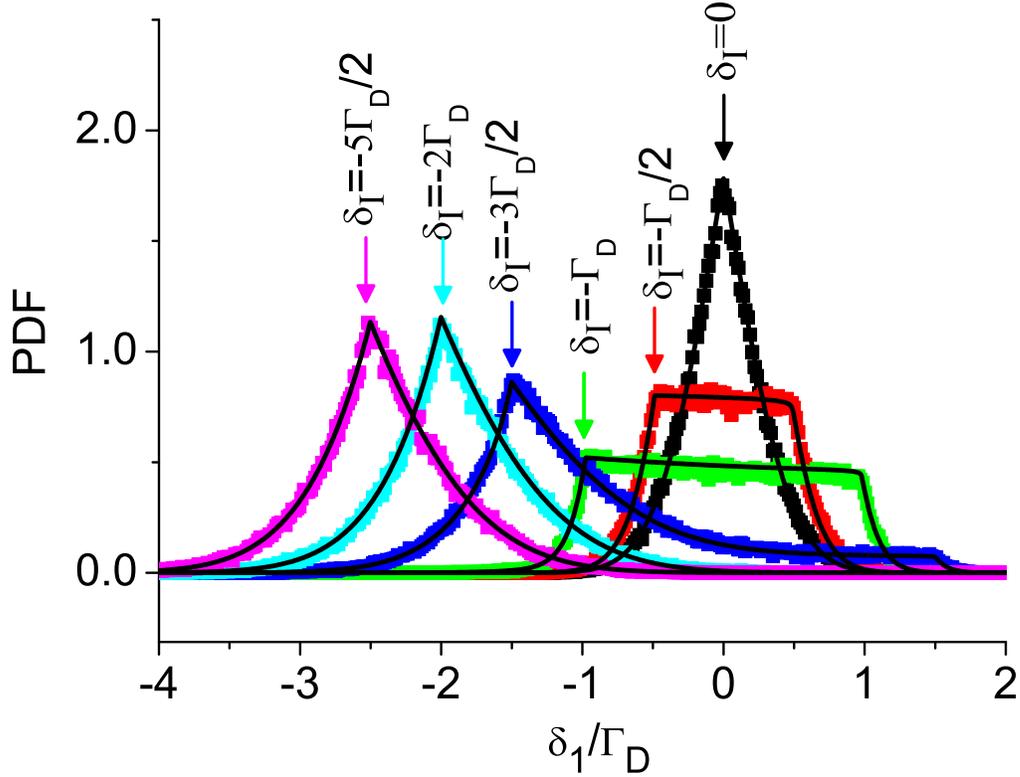}
	\caption{(Color online) Probability density function for the frequency emitted after the first scattering event ($\delta_1$). The arrows in the Figure indicate the excitation frequency ($\delta_I$) for each curve. Calculated $R(\delta_1,\delta_I)$ using equation 3.12.1 given by Hummer \cite{Hummer62} are shown as solid lines.}
	\label{fig:FirstScattering}
\end{figure}

In this section we discuss the results of the simulation of radiation trapping in an \textit{infinite vapor}. We plot in Figure \ref{fig:FirstScattering} the PDF of the photons detuning after the first scattering event. The frequency is measured in the laboratory frame for several initial excitation frequencies ($\delta_I$). One may observe that for small excitation detunings, the reemission occurs around the line center while for excitation far from resonance the emission is centered at the incident frequency. Those curves are equivalent to the redistribution function $R(\delta_1,\delta_I)$ given by Hummer \cite{Hummer62}. Curves calculated using equation 3.12.1 given Hummer \cite{Hummer62} are shown as solid lines in Figure \ref{fig:FirstScattering}.\\

\begin{figure}
	\centering
		\includegraphics[width=1.0\linewidth]{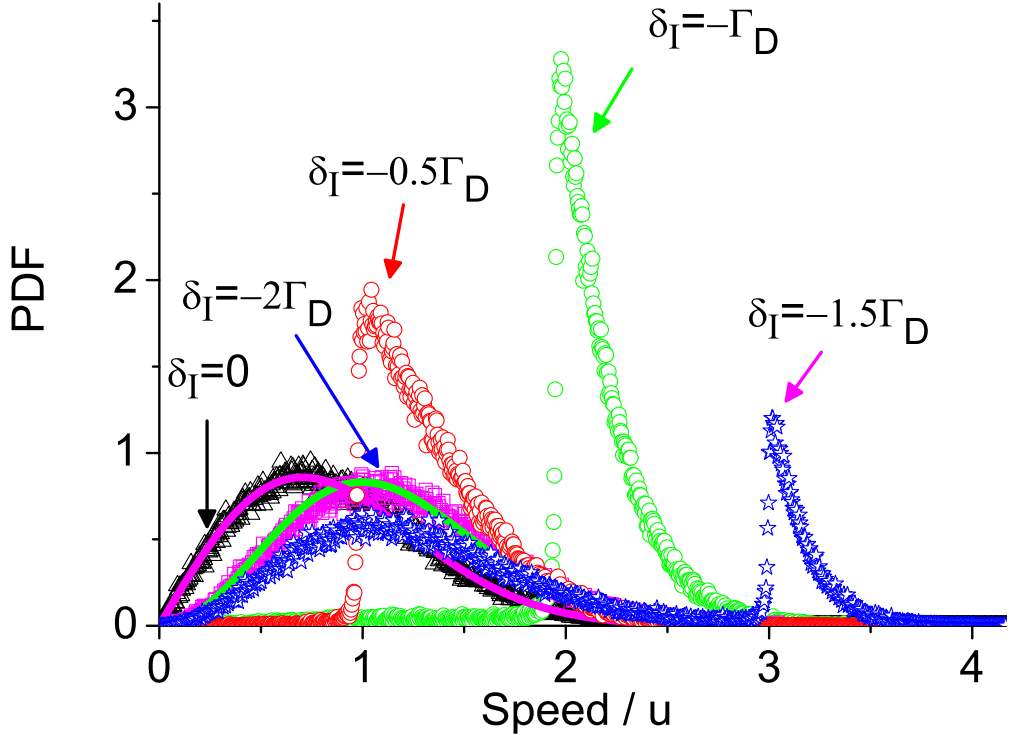}
	\caption{(Color online) Probability density function of the speed of the absorbing atoms, for different excitation detunings $\delta_I$. The magenta solid line superposed to the $\delta_I=0$ curve is a 2D MB distribution at 294 K. The green solid line superposed to the $\delta_I=-2\Gamma_D$ curve is a 3D MB distribution at 294 K.}
	\label{fig:ProbModV}
\end{figure}

The qualitatively different behaviors of ``into resonance" ($\left|\delta_I\right|<\delta_L$) and ``wing" excitation ($\left|\delta_I\right|>\delta_L$) are clearly shown in Figure \ref{fig:FirstScattering}. They are directly related to the two distinct mechanisms of absorption discussed in section \ref{Velocity of the atom absorbing the photon}. For slightly-detuned excitation, the photon is absorbed by an atom with $V_{||}\approx\lambda\delta_I$, i.e., the photon frequency is Doppler-shifted to the line center in the atomic rest frame ($\delta_{A,I}\approx0$). Reemission also occurs at $\delta_{A,1}\approx0$ and the frequency is Doppler-shifted in the laboratory frame. For excitation at $\delta_I=0$, $P_{\delta_I}(V_{||})$ is centered at zero velocity. The contribution of the very narrow distribution of the parallel component to the speed value is negligible and the PDF of the speed of the absorbing atoms approximately follows a 2D Maxwell-Boltzmann distribution (see Figure \ref{fig:ProbModV}).  The most probable speed for the 2D MB distribution is smaller than that for the 3D MB distribution, resulting in a Doppler-shifted reemission narrower than the absorption profile \cite{Voigt} and centered at $\delta_1=0$. For excitation detunings smaller than $\delta_L$ ($\delta_L=1.7\Gamma_D$ in our system), the absorbing atom has a well defined velocity component parallel to the incoming photon $V_{||}\approx\lambda\delta_I$ and a distribution of velocity component normal to the incoming photon centered at zero. The atom speed therefore cannot be smaller than $\left|V_{||}\right|$ (see Figure \ref{fig:ProbModV}) and the acquired Doppler shift in emission is essentially given by  $\left(V_{||}/\lambda\right) (\cos\theta-1) \approx \delta_I (\cos\theta-1)$ (see Eq.\ref{eq:Dshift}). For an isotropic emission, $\cos(\theta)$ is uniformly distributed between $-1$ and $1$, resulting in a plateau-like emission spectrum, of width $\sim2\delta_I$ and amplitude proportional to $1/\delta_I$.\\

For excitations with a detuning larger than $\delta_L$, absorption occurs off resonance in the frame of atoms following a thermal 3D MB distribution (see Figure \ref{fig:ProbModV}). Reemission occurs at the absorbed frequency in the atomic frame ($\delta_{A,I}$ around $\delta_I$). In the laboratory frame the emission frequency distribution is centered at $\delta_I$ and has a Doppler width.

\subsubsection{Many scattering events}
\begin{figure}
	\centering
		\includegraphics[width=1.0\linewidth]{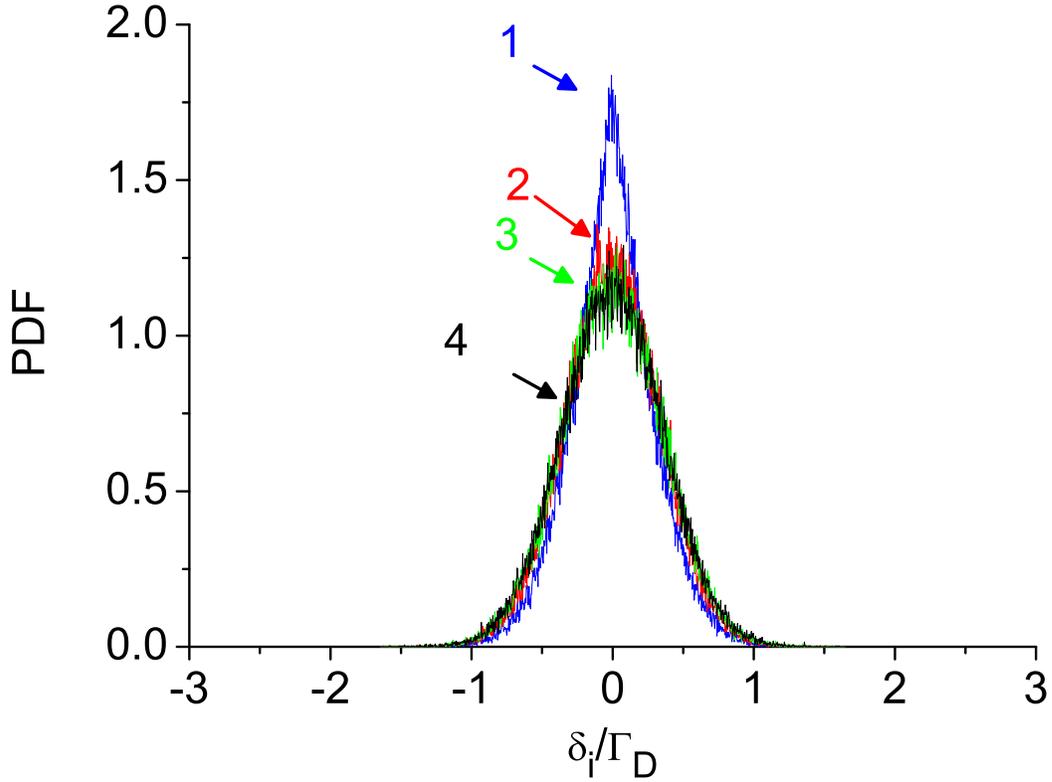}
	\caption{(Color online) Probability density function of the emitted frequency for different scattering event number $i$ (denoted by the numbers in the figure). The initial excitation is at the line center ($\delta_I=0$).}
	\label{fig:line centerInf}
\end{figure}

\begin{figure}
	\centering
		\includegraphics[width=1.0\linewidth]{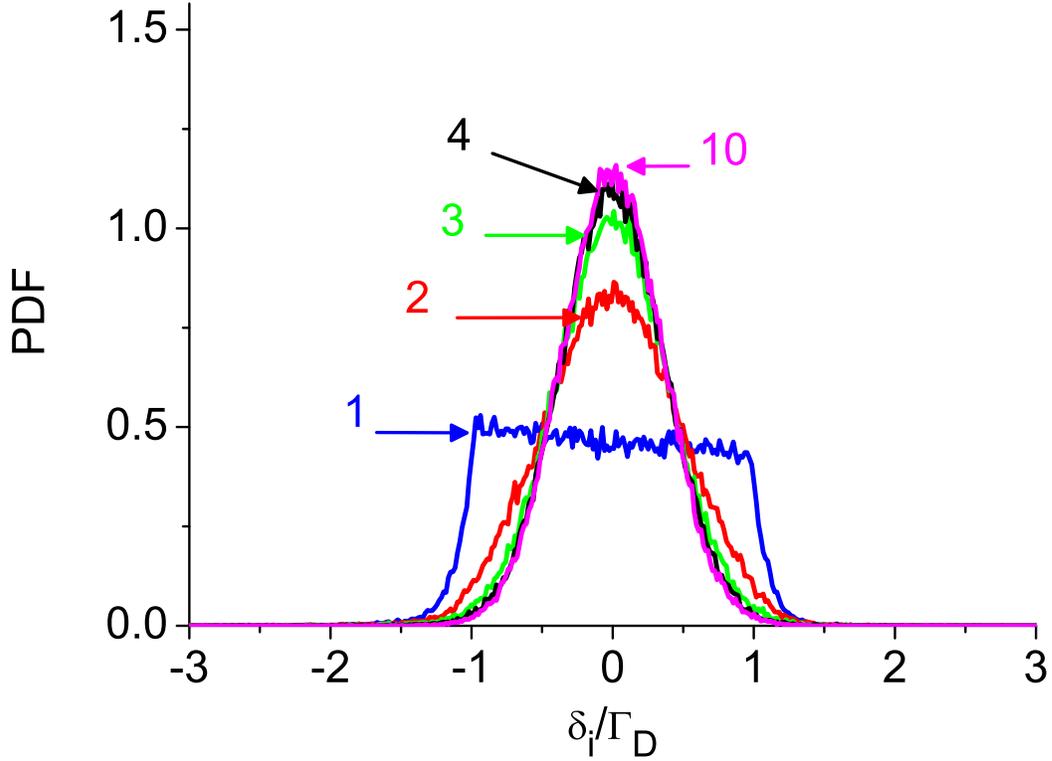}
	\caption{(Color online) Probability density function of the emitted frequency for different scattering event number $i$ (denoted by the numbers in the figure). The initial excitation is $\delta_I=-\Gamma_D$.}
	\label{fig:600MHzInf}
\end{figure}

We now turn to the observation of how the distribution of the emitted frequency evolves with the number of scattering events. In Figure \ref{fig:line centerInf} is shown the PDF for an excitation at line center ($\delta_I=0$) and in Figure \ref{fig:600MHzInf} for excitation at $\delta_I=-\Gamma_D$. For $\delta_I\neq0$ but smaller than $\delta_L$, the first diffusion produces an almost uniform distribution of width $2\delta_I$ around the line center, as discussed in section \ref{sec:freqredis1}. As the emitted frequency after the first scattering obeys $\left|\delta_1\right|<\delta_L$, each ensemble of photons with a given $\delta_1$ will produce after the second scattering a plateau of height $\propto1/\delta_1$ and width $2\delta_1$. This results in a PDF centered at $\delta_2=0$ with width larger than $\Gamma_D$ and smaller than $2\delta_I$. The following scattering events are very close to the absorption profile of the vapor, converging to a CFR situation.\\

\begin{figure}
	\centering
		\includegraphics[width=1.0\linewidth]{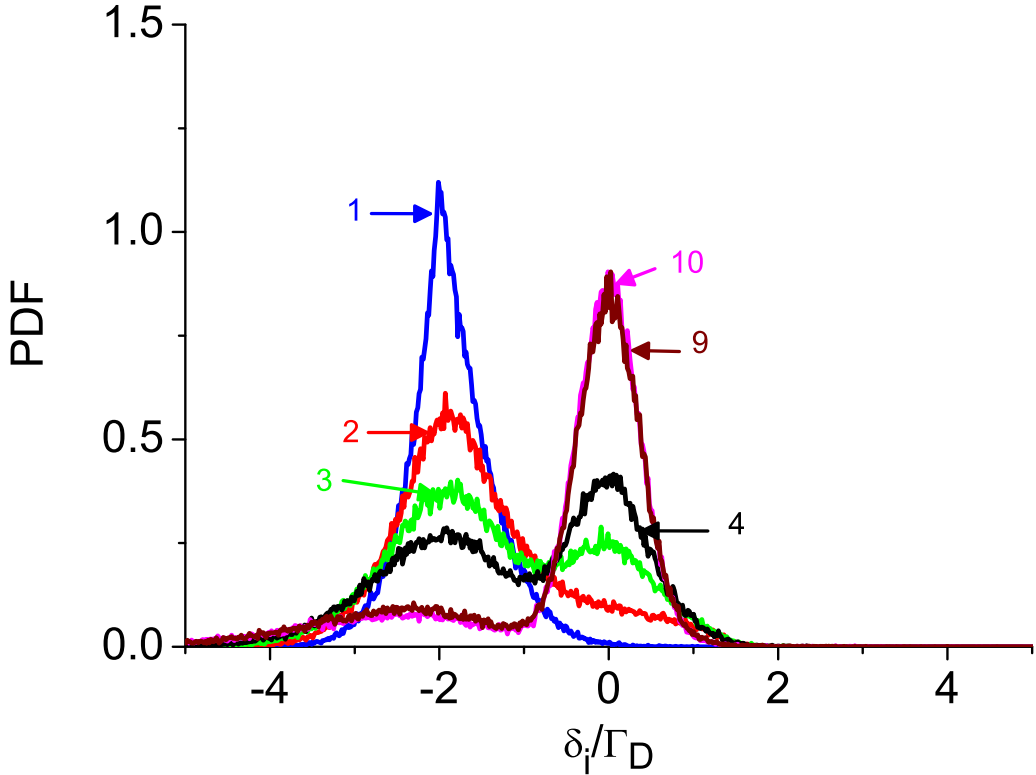}
	\caption{(Color online) Probability density function of the emitted frequency for different scattering event number $i$ (denoted by the numbers in the figure). The initial excitation is at $\delta_I=-2\Gamma_D$.}
	\label{fig:2GHzInf}
\end{figure}

Let us examine the emitted frequency PDF for large excitation detunings ($\delta_I>\delta_L$). The memory of the incident frequency is partially kept after the first scattering (see Figure \ref{fig:2GHzInf}). However, the Doppler broadening of the emission frequency $\delta_1$ around the excitation frequency $\delta_I$ results in some photons close enough to resonance ($\delta_1<\delta_L$) to be subsequently absorbed by atoms with $V_{||}=\lambda\delta_1$. Those photons are thus reemitted with detuning $\delta_2$ around the line center. As a result, for this second scattering, a double-peak appears in the frequency distribution: one peak is centered at $\delta_2=\delta_I$ and the other one centered at $\delta_2=0$. In the following scattering events ($i>1$), the peak around the line center gets higher until it dominates the emission frequency PDF. This behavior results in a net diffusion of the emitted frequency of the whole sample and has consequences for the transport of photons in the vapor.

\begin{figure}
	\centering
		\includegraphics[width=1.0\linewidth]{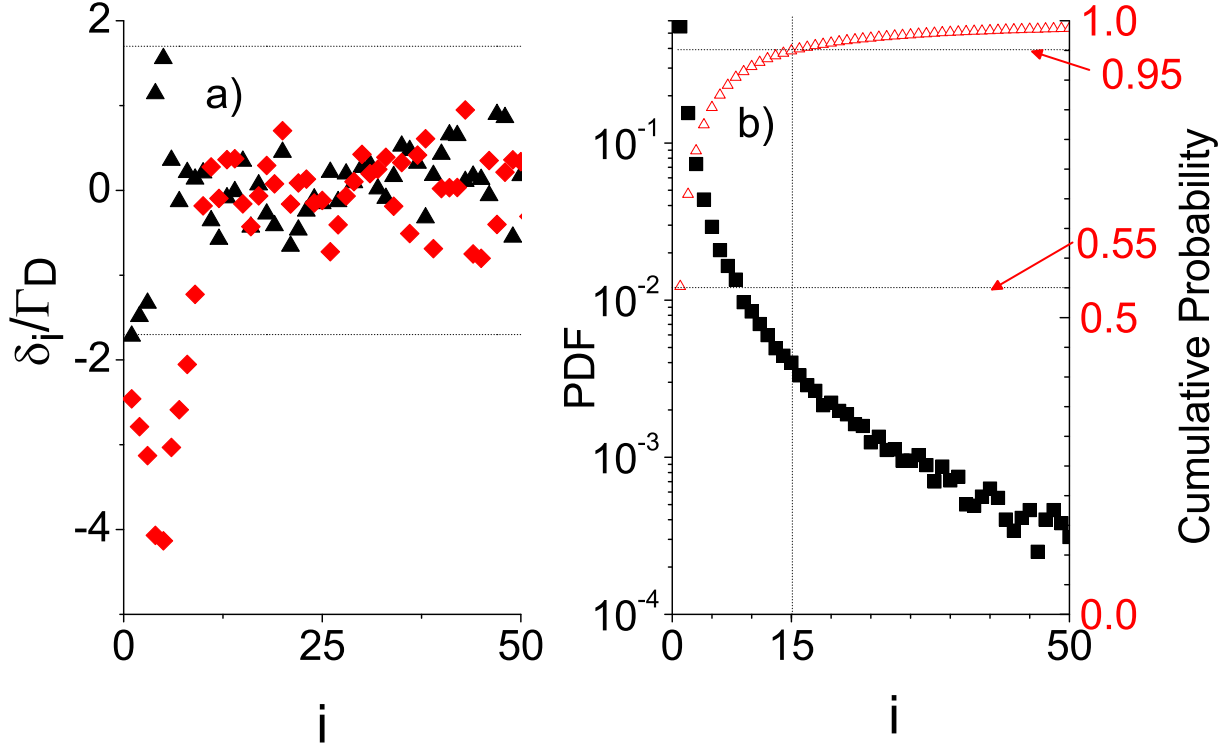}
	\caption{(Color online) a) Frequency diffusion during the random walk of a photon in a vapor as a function of the number of scattering events. Two photon realizations (red diamond and black triangle) are shown for $\delta_I=-2\Gamma_D$. Dotted lines in (a) corresponds to $\pm\delta_L$. b) PDF of the number of scattering events before $\vert\delta_i\vert < \delta_L$ for $\delta_I=-2\Gamma_D$ (left axis) and its cumulative probability (right axis).}
	\label{fig:FreqDif}
\end{figure}

To illustrate the frequency diffusion in the vapor we plot in Figure \ref{fig:FreqDif}a the emission frequency of a photon as a function of the number of scattering events for an excitation detuning $\delta_I=-2\Gamma_D$ for two different realizations of the MC simulation. One may see that at least a few scattering events are necessary for the emitted frequency to be inside the region $\left|\delta_i\right|<\delta_L$ where absorption preferentially happens for an atom with $V_{||}=\lambda\delta_i$. Thereafter, emission occurs around the line center and the emitted frequency profile seems to converge to CFR. We plot in Figure \ref{fig:FreqDif}b the probability density function of the number of events necessary before emission occurs in the region $\left|\delta_i\right|<\delta_L$ for the initial excitation detuning $\delta_I=-2\Gamma_D$ used in Figure \ref{fig:FreqDif}a. For this specific detuning, we see that approximately 55\% of the photons have their frequency in the range $\left|\delta_i\right|<\delta_L$ after the first scatttering. However, a number of 15 scattering events is necessary to put 95\% of the photons in this range.\\

\begin{figure}[htbp]
	\centering
		\includegraphics[width=1.0\linewidth]{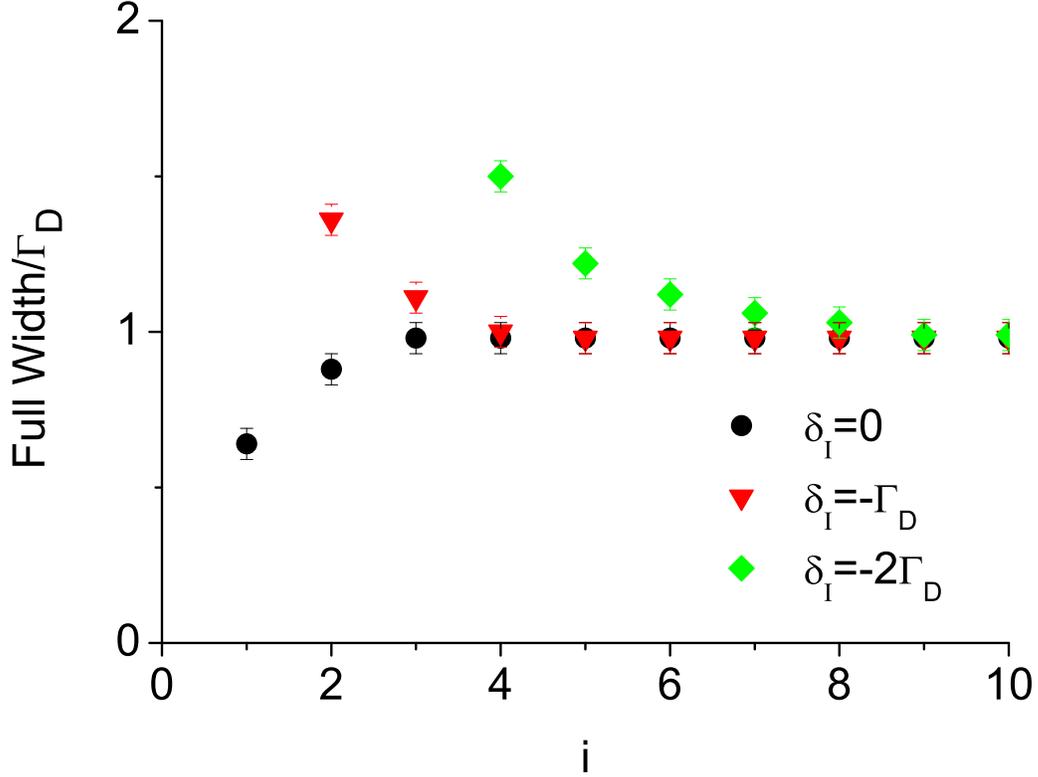}
	\caption{(Color online) Full width at $1/e$ of the maximum of the emitted frequency PDF as a function of the scattering event number for various excitation frequencies $\delta_I$. For $\delta_I=-2\Gamma_D$ only the peak around the line center is reported. For the first three scattering events such peak is not resolved and is not reported here.}
	\label{fig:Swidth}
\end{figure}

The spectral line shape of emission is important to interpret the behavior of the diffusion of photons in a resonant vapor. The frequency redistribution regimes usually considered are PFR after a single diffusion and CFR as the asymptotic result of many diffusion events \cite{Pereira07}. The question that naturally arises is how many cycles of absorption-reemission are necessary to obtain CRF. A way to analyze this question is to monitor the width of the emission spectrum as a function of the number of scattering events, as shown in Figure \ref{fig:Swidth} for different detunings ($\delta_I=0$, $\delta_I=-\Gamma_D$ and $\delta_I=-2\Gamma_D$). We measure the full width at $1/e$ of its maximum in the emission spectra. For $\delta_I=0$ we see that the first scattering produces an emission frequency distribution that is narrower than the absorption spectrum in the vapor in accordance with the discussion in section \ref{sec:freqredis1}. After the third scattering event one can consider that the width does not change much, being equal to $\Gamma_D$. This situation configures a CFR regime. Notice that for the Voigt parameter used ($a=0.01$, $\Gamma_D\gg\Gamma$) the lineshape is dominated by Doppler profile \cite{Voigt} for small detunings and our numerical simulation sampling does not allow to differentiate the Doppler from the Voigt width. For $\delta_I=-\Gamma_D$ the emission spectrum after the first scattering is very broad since it has the plateau-like shape discussed above. For the following scattering events ($i\geq2$) the width gets narrower and reaches a stable value from the fourth scattering event on ($i\geq4$). For $\delta_I=-2\Gamma_D$ we have measured the width of the peak around the line center. For the first three scattering events this peak is not resolved and is not reported. We see that for small numbers of events the frequency distribution is broad and gets narrower until its width stabilizes at $\Gamma_D$, around the ninth scattering event. 

\subsection{Photon step-length distribution}
\begin{figure}
	\centering
		\includegraphics[width=1.0\linewidth]{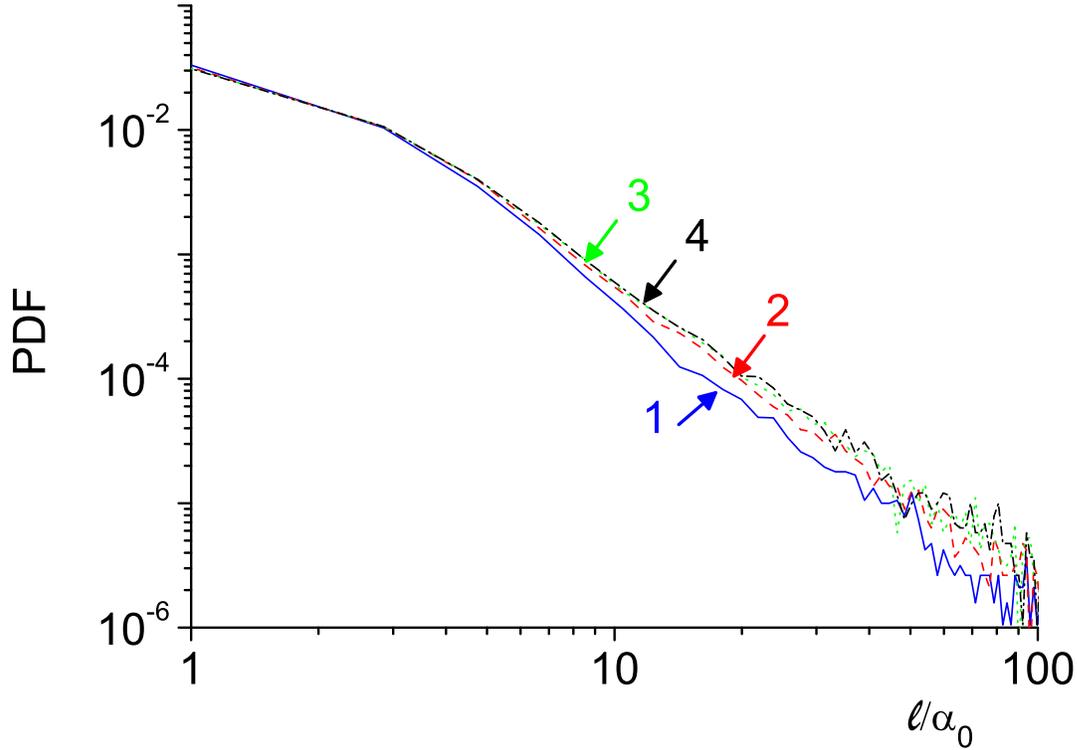}
	\caption{(Color online) Probability density function of the step length ($P(l)$) of the photon before being reabsorbed by the vapor for different scattering event number i (denoted as numbers in the figure). Excitation is at the line center $\delta_I=0$. The step length is normalized by the absorption coefficient at the line center $\alpha_0$.}
	\label{fig:Pl0}
\end{figure}

\begin{figure}
	\centering
		\includegraphics[width=1.0\linewidth]{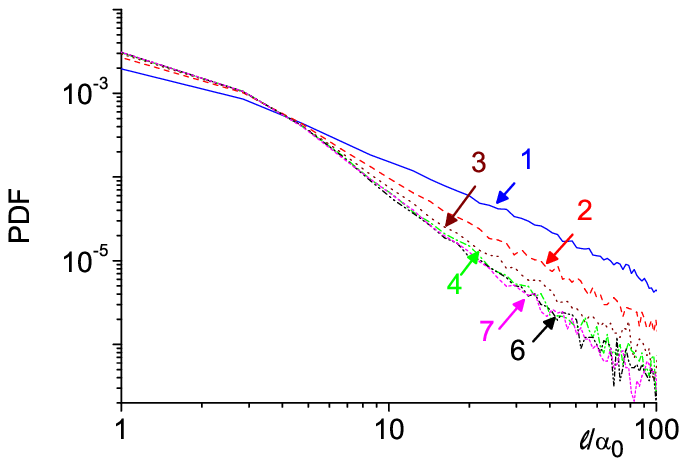}
	\caption{(Color online) Probability density function of the step length ($P(l)$) of the photon before being reabsorbed by the vapor for different scattering event number i (denoted as numbers in the figure). Excitation is at $\delta_I=-\Gamma_D$. The step length is normalized by the absorption coefficient at the line center $\alpha_0$.}
	\label{fig:Pl600}
\end{figure}

\begin{figure}
	\centering
		\includegraphics[width=1.0\linewidth]{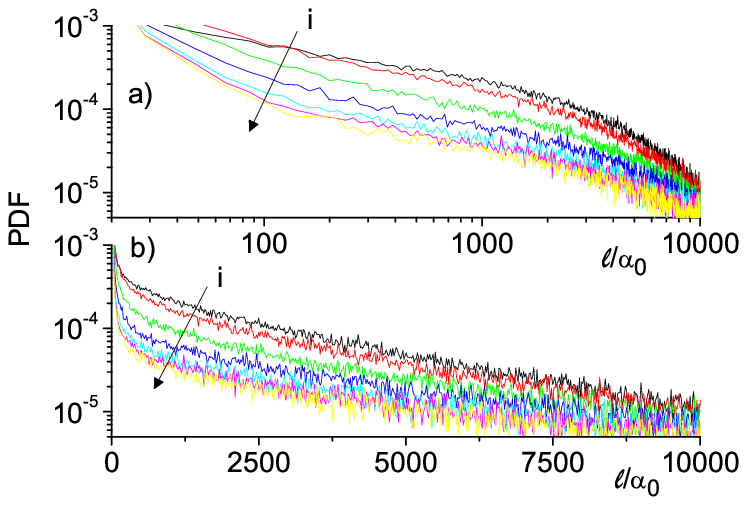}
	\caption{(Color online) Probability density function of the step length ($P(l)$) of the photon before being reabsorbed by the vapor for different scattering event number i. The arrow indicate the sequence of curves with $i=1,2,4,6,8,9,10$. Excitation is at $\delta_I=-2\Gamma_D$. The step length is normalized by the absorption coefficient at the line center $\alpha_0$.}
	\label{fig:Pl4GD}
\end{figure}

The wings of the emission and of the absorption spectral distributions play a crucial role in the transport of radiation through a vapor, since a photon emitted in the spectrum wings, i.e. far from resonance center, travels a longer path before being absorbed than if it is emitted close to the center of the resonance \cite{Kenty32}. In particular, the asymptotic behavior of the step-length PDF ($P(l)$) dictates the statistics of the diffusion \cite{Pereira04,Pereira07}. Measurements of $P(l)$ in the CFR configuration require the preparation of a radiation with spectral profile identical to the absorption one, which can be achieved, e.g., through multiple scattering of laser radiation inside an atomic vapor prior to sending these photons to the measurement cell \cite{NatPhys,Mercadier13}. Notice, however, that to determine if CFR is achieved, the criterion of stability of the emitted spectrum width is not the most adequate since it gives little information on the spectrum wings. We show in Figures (\ref{fig:Pl0}) to (\ref{fig:Pl4GD}) the evolution of $P(l)$ as a function of the number of scattering events for the three values of $\delta_I$ used in Figures \ref{fig:line centerInf}, \ref{fig:600MHzInf} and  \ref{fig:2GHzInf}. For $\delta_I=0$ (Figure \ref{fig:Pl0}) we see the stability of $P(l)$ from the third scattering event on, indicating that three scattering events are enough to consider that the CFR regime is reached. For $\delta_I=-\Gamma_D$ (Figure \ref{fig:Pl600}), subtle changes of $P(l)$ occur until the PDF stabilizes from the sixth scattering event on. For $\delta_I=-2\Gamma_D$ (Figure \ref{fig:Pl4GD}), at least nine scattering events are needed to achieve a stable $P(l)$.\\

For $\delta_I=0$ (Figure \ref{fig:Pl0}), $P(l)$ after the first scattering decays faster than $l^{-1.5}$, expected for a diffusion of radiation with a Voigt profile into a Voigt absorption medium \cite{Pereira07}. This is due to the fact that the emission spectrum after the first scattering is narrower than the Voigt lineshape and thus closer to a monochromatic spectrum. For $\delta_I=-\Gamma_D$ (Figure \ref{fig:Pl600}), $P(l)$ stabilizes with an asymptotic power law $l^{-2}$ typical of a Doppler radiation incident in a vapor with Doppler absorption profile. This is expected for CFR in a vapor with Voigt absorption profile \cite{Chevrollier,Pereira04} with Voigt coefficient $a=0.01$ in the range of $l/\alpha_0$ exhibited in Figure \ref {fig:Pl600}. Our MC simulation has not reached the asymptotic range $P(l)\propto l^{-1.5}$ ($l/\alpha_0>10^5$) corresponding to very rare events far in the wings of the spectral distribution. For $\delta_I=-2\Gamma_D$ (Figure \ref{fig:Pl4GD}), $P(l)$ exhibits a power-like law for intermediate values of step length (see linear part of the curve in a log-log scale in Figure \ref{fig:Pl4GD}a). For the asymptotic long steps, $P(l)$ behaves rather as an exponential decay (see linear part of the curve in a mono-log scale in Figure \ref{fig:Pl4GD}b). The long steps are taken by photons emitted far from resonance in the broad peak around $\delta_i=\delta_I=-2\Gamma_D$, i.e., in the wings of the absorption profile. As the Voigt absorption profile falls as $\delta^{-2}$, the absorption coefficient around $-2\Gamma_D$ changes slowly in the far wings and $P(l)$ approaches a behavior described by a Beer-Lambert law. This corresponds to a change in the statistical behavior of the light transport from superdifusive (for $\left|\delta_I\right|<\delta_L$) to normal diffusion (for $\left|\delta_I\right|>\delta_L$) \cite{Pereira07}. The emission peak in the wing of the resonance around $\delta_i=\delta_I=-2\Gamma_D$ is broadened with the number of scattering events (see Figure \ref{fig:2GHzInf}). As a result, for large numbers of scattering events, an almost constant emission profile appears in the wing of the resonance line. The consequence of the subsistence of this constant profile is that CFR is never reached, that is, the emission profile does not completely converge to the Voigt absorption profile and $P(l)$ maintains its exponential-like asymptotic regime over large numbers of scattering events.

\section{Conclusion}
We have developed a Monte-Carlo simulation to analyze the diffusion of light propagating in a resonant atomic vapor. For the first scattering event, our simulations give results similar to the theoretical approach of the redistribution function developed by Unno \cite{Unno52} and Hummer \cite{Hummer62}. We have obtained the evolution of the PDF of the emitted frequency after multiple scattering events and analysed its convergence to a Complete Frequency Redistribution. Using only the width of the emitted spectrum as a criterion to check CFR, we observe that an average of three scattering events is necessary to reach CFR when the excitation is tuned to the line center. On the other hand, an average of six scattering events is necessary to fully redistribute an excitation detuned one Doppler width away from the resonance; Nine scattering events are necessary for excitation at two Doppler widths from resonance. A second criterion can be used to analyse CFR: the Probability Density Function of the step length of photons between an emission and the subsequent absorption in the vapor, $P(l)$. For excitation close to resonance $\left|\delta_I\right|<\delta_L$, the long tail of $P(l)$ behaves as a power law that changes its coefficient at each scattering event. For those frequencies the photons undergo a superdiffusive transport inside the vapor. However, for excitation far from resonance ($\left|\delta_I\right|>\delta_L$) $P(l)$ exhibits an exponential decay typical of normal diffusive transport. The exponential behavior originates from the photons emitted around the excitation frequency. Such a far-from-resonance emission subsists for a large number of scattering events preventing CFR to be reached and the transport maintains normal diffusion characteristics. We believe that these MC simulations allow to better interpret data from experiments of light diffusion in a resonant vapor. A natural prospect of this work is to include in the simulations the multilevel structure of the Rb isotopes. Indeed, optical pumping can occur between different ground states of an alkali atom and, moreover, transfer of radiation can occur between the two Rb isotopes due, e.q., to the proximity of the levels $^{85}5S_{1/2}(F=3)$ and $\:^{87}5S_{1/2}(F=2)$.\\

\begin{acknowledgments}
We thank financial support from Brazilian agencies CNPq, FINEP, and CAPES. 
\end{acknowledgments}

\end{document}